\documentclass[aps,prx,reprint,superscriptaddress,floatfix]{revtex4-2}
\usepackage{amssymb,amsfonts,dsfont,amsmath}
\usepackage{times} 
\usepackage{graphicx} 
\usepackage{bm}
\usepackage{accents}
\usepackage{mathtools}
\usepackage[colorlinks=true,linktocpage=true,breaklinks=true]{hyperref}
\usepackage{xcolor}

\newcommand{\be}{\begin{equation}} 
\newcommand{\e}{\end{equation}} 
\newcommand{\bb}{\boldsymbol}

\DeclareMathOperator{\tr}{Tr} 
\DeclareMathOperator{\mtr}{tr}

\begin{document}

\title{Kubo formula for dc conductivity: generalization to systems with spin-orbit coupling}

\author{I.\,A.~Ado}
\email{iv.a.ado@gmail.com}
\affiliation{Radboud University, Institute for Molecules and Materials, 6525 AJ Nijmegen, The Netherlands}
\affiliation{Institute for Theoretical Physics, Utrecht University, 3584 CC Utrecht, The Netherlands}

\author{M.\,Titov}
\affiliation{Radboud University, Institute for Molecules and Materials, 6525 AJ Nijmegen, The Netherlands}

\author{Rembert A.\,Duine}
\affiliation{Institute for Theoretical Physics, Utrecht University, 3584 CC Utrecht, The Netherlands}
\affiliation{Department of Applied Physics, Eindhoven University of Technology, P.O. Box 513, 5600 MB Eindhoven, The Netherlands}

\author{Arne Brataas}
\affiliation{Center for Quantum Spintronics, Department of Physics, Norwegian University of Science and Technology, H\o gskoleringen 5, 7491 Trondheim, Norway}

\begin{abstract}
We revise the Kubo formula for the electric dc conductivity in the presence of spin-orbit coupling (SOC). We discover that each velocity operator that enters this formula differs from $\partial H/\partial \bb p$, where $H$ is the Hamiltonian and $\bb p$ is the canonical momentum. Moreover, we find an additional contribution to the Hall dc conductivity from noncommuting coordinates that is missing in the conventional Kubo-St\v reda formula. This contribution originates from the ``electron-positron'' matrix elements of the velocity and position operators. We argue that the widely used Rashba model does in fact provide a finite anomalous Hall dc conductivity in the metallic regime (in the noncrossing approximation) if SOC-corrections to the velocity and position operators are properly taken into account. While we focus on the response of the charge current to the electric field, linear response theories of other SOC-related effects should be modified similarly.
\end{abstract}

\maketitle
\section{Introduction}
Approximately 25 years ago, out-of-equilibrium phenomena caused by spin-orbit coupling (SOC) started gaining strong attention. Among such phenomena are the anomalous Hall effect (AHE)~\cite{RevModPhys.82.1539}, the spin Hall effect (SHE)~\cite{RevModPhys.87.1213}, the spin-orbit torques (SOT)~\cite{RevModPhys.91.035004}, and their thermoelectric and spin caloritronics counterparts~\cite{bauer2012spin}. These effects continue to be the subject of very active and fruitful research today. Many ideas for technological applications based on them have been suggested and developed, such as SOT magnetoresistive random-access memory~\cite{8502269,SAHA2022169161}, magnetic racetrack memory~\cite{parkin2015memory}, and neuromorphic computing~\cite{grollier2020neuromorphic}. AHE is often considered as a hallmark of the nontrivial topology of the system~\cite{RevModPhys.82.1959}. It also serves as an important probe of magnetic order~\cite{bonilla2018strong, tan2018hard, deng2018gate, alghamdi2019highly}.

From the computational point of view, these effects can be analyzed using kinetic-type equations or different versions of the linear response Kubo formula~\cite{kubo1957statistical}. The latter approach normally allows one to compute the corresponding quantities in the most direct and reliable way. In this paper, however, we show that, for systems with SOC, the existing Kubo formulas miss important contributions and should be revised. One of the reasons for this lies in the fact that the unitary transformation that diagonalizes a relativistic Hamiltonian does not, in general, diagonalize other operators of physical observables. While the spectrum of the system and its ground state properties are not affected by this, linear response theories are. Another important issue is that the operator $\partial/\partial\bb p$ is not invariant under $\bb p$-dependent unitary transformations. Below, using the electrical dc conductivity as an example, we demonstrate that a consistent treatment of observables and perturbations reveals previously overlooked contributions to out-of-equilibrium phenomena originating from SOC.

In the nonrelativistic (nr) theory, the Kubo formula for the full dc conductivity tensor of an electron gas was derived in 1971 by Bastin \textit{et al.}, using the Green's function formalism~\cite{BASTIN19711811}. It can be cast in the form
\begin{equation}
    \label{nr_Bastin}
    \sigma^{\text{nr}}_{\alpha\beta}=\frac{e^2\hbar}{2\pi\Omega}\tr\int d\varepsilon\,f_\varepsilon(G_--G_+) v_\alpha G_+' v_\beta+\text{c.c.},
\end{equation}
where $G_{\pm}=(\varepsilon\pm i 0-H)^{-1}$ denotes the retarded (+) and advanced (-) Green's functions, $H$ is the Hamiltonian, $f_\varepsilon$ is the Fermi-Dirac distribution function, $\Omega$ represents the system volume, $\bb v=[\bb r, H]/i\hbar=\partial H/\partial \bb p$ is the velocity operator, $\bb p$~is the canonical momentum, $\bb r$ represents the position operator, and 
prime denotes a derivative with respect to $\varepsilon$. The system is assumed to be noninteracting.

In 1982, St\v reda published a modified version of the Kubo-Bastin formula, in which he split the nonrelativistic conductivity tensor into a sum of two distinct contributions~\cite{PStreda_1982}, $\sigma_{\alpha\beta}^{\text{nr}}=\sigma^{\text{nr-I}}_{\alpha\beta}+\sigma^{\text{nr-II}}_{\alpha\beta}$. By applying a partial integration to Eq.~(\ref{nr_Bastin}) and then taking a half-sum of the obtained result and the original formula, one can express the St\v reda's splitting as
\begin{subequations}
\label{nr_Streda}
\begin{gather}
\label{Streda_1}
\sigma^{\text{nr-I}}_{\alpha\beta}=\frac{e^2\hbar}{4\pi\Omega}\tr\int d\varepsilon\,f'_\varepsilon(G_+-G_-) v_\alpha G_+ v_\beta+\text{c.c.},
\\
\begin{aligned}
\label{Streda_2}
\sigma^{\text{nr-II}}_{\alpha\beta}=\frac{e^2\hbar}{4\pi\Omega}\tr\int d\varepsilon\,f_\varepsilon
(&G_- v_\alpha G_- v_\beta G_-\\ &- G_- v_\beta G_- v_\alpha G_-)+\text{c.c.}
\end{aligned} 
\end{gather}
\end{subequations}

At the end of the 1990s, when the interest in SOC-related out-of-equilibrium phenomena increased, St\v reda's formula [Eqs.~(\ref{nr_Streda})] and its modifications became standard tools for computing tensors that describe such effects. For example, to obtain the dc spin Hall conductivity, one would replace the velocity operator $v_\alpha$ in Eqs.~(\ref{nr_Streda}) with the spin velocity operator. For SOT, $v_\alpha$ would change to the spin operator. For AHE, the St\v reda's formula was used directly.

In the vast majority of all such computations, the velocity operators are defined as $\bb v=\partial H/\partial\bb p$. This is because the commutation of the Hamiltonian with $\bb r$ corresponds to differentiating the former with respect to $\bb p$. For AHE, this is also in line with the conventional definition of the electric current density operator, which is considered proportional to the functional derivative $\delta H/\delta \bb A$. Here $\bb A$ is the vector potential and, according to the Peierls substitution, it enters the Hamiltonian through the combination $\bb \pi=\bb p-(e/c)\bb A$~\footnote{For a system described by the Pauli Hamiltonian, $\bb A$ also contributes to the Zeeman term through $\bb B=\bb\nabla\times\bb A$, but its derivative with respect to $\bb A$ does not contribute to the transport current.}.

At the same time, it is known that the position operator $\bb r$ acquires a relativistic (or effectively relativistic) correction $\Delta\bb{r}$ both in vacuum~\cite{PhysRev.78.29} and in crystals~\cite{nozieres1973simple, vignale2010ten}. This correction is sometimes called the Yafet term~\cite{kronmuller2007handbook, PhysRevB.95.054509}. Upon commuting the modified coordinate operator with the Hamiltonian, one finds that the velocity operator should be corrected as well,
\begin{equation}
\bb v=\frac{[\bb r+\Delta\bb r,H]}{i\hbar}=\frac{\partial H}{\partial \bb p}+\bb v_{\text{an}},
\end{equation}
where $\bb v_{\text{an}}=[\Delta\bb r, H]/i\hbar$ is what Adams and Blount called the ``anomalous velocity'' in 1959~\cite{comment, adams1959energy, BLOUNT1962305}. Sometimes it is confused with the spin-dependent part of the velocity operator.

Now we can formulate the problem. On the one hand, we want to use Kubo formulas for effects like AHE, SHE, and SOT, which are relativistic (as they originate from SOC). On the other hand, Eqs.~(\ref{nr_Streda}) were derived for a nonrelativistic system. A priori, it is unclear how the anomalous velocity $\bb v_{\text{an}}$ and the Yafet term $\Delta\bb{r}$ enter Kubo formulas in the presence of SOC. Thus, a separate derivation of the Kubo formulas is required in this case to include all effects of SOC consistently.

For the conductivity tensor $\sigma_{\alpha\beta}$, this task was approached by Cr\'epieux and Bruno in 2001. They found an elegant expression for the Green's function of the Dirac Hamiltonian in terms of the nonrelativistic Green's function~\cite{PhysRevB.64.094434}. In Ref.~\cite{PhysRevB.64.014416}, they used this expression to project the St\v reda's formula derived in the Dirac theory onto the electronic branch. This was done by expanding the latter formula up to the order $1/c^2$, where $c$ is the speed of light. According to the result of Cr\'epieux and Bruno, Eqs.~(\ref{nr_Streda}) with $\bb v=\partial H/\partial \bb p$ still hold when the leading order relativistic corrections, including SOC, are taken into account. In other words, the anomalous velocity $\bb v_{\text{an}}$ is absent in their formulas. The origin of this is that the Yafet term is missing in the definition of the velocity operator (see, e.g., Eqs.~(28) and (30) of Ref.~\cite{PhysRevB.64.094434}). In 2015, another work~\cite{PhysRevB.91.174415} extended some of the results of Cr\'epieux and Bruno, but also disregarded the Yafet term.

Below, we derive the St\v reda's formula for electrons in the presence of SOC, starting from the Dirac Hamiltonian. We use a different method of projecting the relativistic Kubo formula onto the electronic branch. Our approach can be used to reach any required accuracy with respect to the parameter $1/c$ in a unified fashion. We find, in particular, that each velocity operator that enters the St\v reda's formula for electrons contains the anomalous part. This applies not only to the velocity operator corresponding to the computed observable but also to the velocity operator representing the Hamiltonian's perturbation by the electric field.

The important principle that we use is that the unitary transformation applied to the Dirac Hamiltonian to remove positrons should also be applied to all other operators. In particular, SOC corrections must be computed not only for effective Hamiltonians, but for observables and perturbations too. Moreover, we show that a proper linear response theory for a system with SOC should also take into account the ``electron-positron'' matrix elements of the involved operators. While the former principle was already formulated, e.g., in Ref.~\cite{Chang_2008}, the latter one, to the best of our knowledge, was not discussed before. We note that while our analysis focuses on the conductivity tensor, our approach can be applied to other SOC-related effects. In what follows, we also discuss several implications of the obtained results.

\section{Unitary transformation of the Dirac picture operators}
Let us outline how, by a unitary transformation, operators of the Dirac theory are projected onto the electronic branch. The Dirac (D) Hamiltonian reads
\begin{equation}
H^{\text{(D)}}=
\begin{pmatrix}
m c^2 + V & c\,\bb \sigma\bb \pi \\
c\,\bb \sigma\bb \pi & -m c^2 + V
\end{pmatrix}
=H^{\text{(D)}}_{\text{d}}+H^{\text{(D)}}_{\text{od}},
\end{equation}
where 
$\bb\sigma$ denotes Pauli matrices, $\bb \pi=-i\hbar \bb\nabla-(e/c)\bb A$,
\begin{equation}
H^{\text{(D)}}_{\text{d}}=
\begin{pmatrix}
H^{\text{(D)}}_+ & 0 \\
0 & H^{\text{(D)}}_-
\end{pmatrix}, \quad H^{\text{(D)}}_{\pm}=\pm mc^2 +V,
\end{equation}
is the diagonal (d) component of the Hamiltonian, while
\begin{equation}
\label{h}
H^{\text{(D)}}_{\text{od}}=
\begin{pmatrix}
0 & h \\
h & 0
\end{pmatrix},\quad h=c\,\bb \sigma\bb \pi,
\end{equation}
is the off-diagonal (od) one. We want to diagonalize $H^{\text{(D)}}$ up to a desired order in $1/c$ by applying a unitary transformation. Such a procedure is sometimes called the L\"owdin partitioning~\cite{winkler2003spin,sinova2008theory}.

We define a unitary operator $U=e^{iW}$ with the help of
\be
W=\begin{pmatrix}
0 & w \\
w^\dagger & 0
\end{pmatrix},
\e
where $w^\dagger$ is a Hermitian conjugate of $w$. Diagonal and off-diagonal components of the unitary transformed Dirac Hamiltonian $H^{\text{(D)}}_U=U^{\dagger} H^{\text{(D)}} U$ are then represented by~\cite{winkler2003spin}
\begin{subequations}
\label{Hu}
\begin{gather}
\label{Huo}
H^{\text{(D)}}_{U,\text{d}}=H^{\text{(D)}}_{\text{d}}+\sum_{k=1}^{\infty}\left[\frac{[H^{\text{(D)}}_{\text{od}},iW]^{(2k-1)}}{(2k-1)!}+\frac{[H^{\text{(D)}}_{\text{d}},iW]^{(2k)}}{(2k)!}\right],\\
\label{Huod}
H^{\text{(D)}}_{U,\text{od}}=H^{\text{(D)}}_{\text{od}}+\sum_{k=1}^{\infty}\left[\frac{[H^{\text{(D)}}_{\text{d}},iW]^{(2k-1)}}{(2k-1)!}+\frac{[H^{\text{(D)}}_{\text{od}},iW]^{(2k)}}{(2k)!}\right],
\end{gather}
\end{subequations}
where the notation
\begin{equation}
[A,B]^{(n)}=[A,\underbrace{B],B],\dots B]}_{n\text{ times}}
\end{equation}
has been introduced.

We need to solve the equation $H^{\text{(D)}}_{U,\text{od}}=0$ for $w$. We begin by considering only the first two terms on the right hand side of Eq.~(\ref{Huod}). This gives $H^{\text{(D)}}_{\text{od}}+[H^{\text{(D)}}_{\text{d}},iW]=0$, which implies
\begin{equation}
\label{w}
h+i(H^{\text{(D)}}_+ - H^{\text{(D)}}_-)w-i[w,H^{\text{(D)}}_-]=0.
\end{equation}
The difference $H^{\text{(D)}}_+ - H^{\text{(D)}}_-= 2m c^2$ is a large parameter and we can solve Eq.~(\ref{w}) iteratively. The second iteration gives
\be
w=\frac{i\bb\sigma\bb\pi}{2mc}+\frac{\hbar}{4m^2c^3}(\bb\sigma\bb\nabla) V+\dots
\e
However, this is not yet the full result for $w$ up to the order $1/c^3$. We need to consider two more terms of the expansion~(\ref{Huod}). By doing so, we find
\be
\label{w3}
w=\frac{i\bb\sigma\bb\pi}{2mc}+\frac{\hbar}{4m^2c^3}(\bb\sigma\bb\nabla) V-\frac{i(\bb\sigma\bb\pi)^3}{6m^3c^3}+\dots
\e
All other contributions to $w$ are at least of the order $1/c^5$.

Substituting Eq.~(\ref{w3}) into Eqs.~(\ref{Hu}), we deduce that the unitary transformed Dirac Hamiltonian becomes
\be
\label{HeHp}
H^{\text{(D)}}_U=
\begin{pmatrix}
H^{\text{(el)}} & 0 \\
0 & H^{\text{(p)}}
\end{pmatrix},
\e
where $H^{\text{(el)}}$ and $H^{\text{(p)}}$ are determined by Eq.~(\ref{Huo}). We see that the off-diagonal elements of the Hamiltonian $H^{\text{(D)}}$ have been eliminated up to the order $1/c$. Consequently,  $H^{\text{(el)}}$ and $H^{\text{(p)}}$, which are the Hamiltonians of electrons (el) and positrons (p), respectively, are decoupled from each other with the corresponding accuracy. Such a decoupling procedure can be performed up to any order of $1/c$.

Importantly, not just the Hamiltonian $H^{\text{(D)}}$, but all other operators transform similarly. However, the essential difference is that, by eliminating the off-diagonal elements of the Hamiltonian $H^{\text{(D)}}$, we produce (or modify) off-diagonal (electron-positron) terms in other operators. \vspace{-0.15ex}For position and velocity operators, $\bb r^{\text{(D)}}_U=U^{\dagger} \bb r^{\text{(D)}} U$, $\bb v^{\text{(D)}}_U=U^{\dagger} \bb v^{\text{(D)}} U$, we will denote the off-diagonal terms as $\delta\bb r$ and $\delta\bb v$,
\be
\label{rv}
\bb r^{\text{(D)}}_U=
\begin{pmatrix}
\bb r^{\text{(el)}} & \delta\bb r \\
\delta\bb r^\dagger & \bb r^{\text{(p)}}
\end{pmatrix},\quad
\bb v^{\text{(D)}}_U=
\begin{pmatrix}
\bb v^{\text{(el)}} & \delta\bb v \\
\delta\bb v^\dagger & \bb v^{\text{(p)}}
\end{pmatrix}.
\e

It is straightforward to relate the elements of $\bb r^{\text{(D)}}_U$ and $\bb v^{\text{(D)}}_U$ to each other. Since, in the Dirac theory, velocity is a commutator $\bb v^{\text{(D)}}=[\bb r^{\text{(D)}},H^{\text{(D)}}]/i\hbar$ (note that $\bb r^{\text{(D)}}=\bb r$) and since unitary transformations preserve commutators, we observe that
\be
\bb v^{\text{(D)}}_U=\frac{1}{i\hbar}
\begin{pmatrix}
[\bb r^{\text{(el)}},H^{\text{(el)}}] & \delta\bb r H^{\text{(p)}}-H^{\text{(el)}}\delta\bb r \\
\delta\bb r^\dagger H^{\text{(el)}}-H^{\text{(p)}}\delta\bb r^\dagger & [\bb r^{\text{(p)}},H^{\text{(p)}}]
\end{pmatrix}.
\label{vDU}
\e
Hence, for electrons, a familiar relation between velocity and position, $\bb v^{\text{(el)}}\!=\![\bb r^{\text{(el)}},H^{\text{(el)}}]/i\hbar$, holds. In addition, we have derived an important result: $\delta\bb v=(\delta\bb r H^{\text{(p)}}-H^{\text{(el)}}\delta\bb r)/i\hbar$. We will use it below.

Let us also give the explicit results for the velocity and position operators up to the order $1/c^2$,
\begin{subequations}
\label{rv_Pauli}
\begin{gather}
    \bb r^{\text{(el)}}=\bb r+\frac{\hbar}{4m^2 c^2}[\bb\pi\times\bb\sigma],\\
    \label{ve}
    \bb v^{\text{(el)}}=\frac{\bb \pi}{m}-\frac{\hbar}{2m^2 c^2}[\bb\nabla V\times\bb\sigma]-\frac{\bb\pi^2\bb\pi}{2m^3c^2}.
\end{gather}
\end{subequations}
We clearly see that the velocity operator $\bb v^{\text{(el)}}$ differs from the $\bb p$-derivative (or $\bb\pi$-derivative) of the electron Hamiltonian~\footnote{In Eq.~(\ref{He}), we disregard all terms beyond the order $1/c^2$.}
\begin{multline}
\label{He}
H^{\text{(el)}}=mc^2+V+\frac{\bb\pi^2}{2m}-\frac{e\hbar}{2mc}\bb\sigma\bb B-\frac{\bb\pi^4}{8m^3c^2}\\+
\frac{\hbar^2}{8m^2c^2}\bb\nabla^2V+\frac{\hbar}{4m^2c^2}[\bb\nabla V\times\bb\pi]\bb\sigma.
\end{multline}
The difference between the velocity operator of Eq.~(\ref{ve}) and $\partial H^{\text{(el)}}/\partial\bb p$ equals $(-\hbar/4m^2 c^2)[\bb\nabla V\times\bb\sigma]$. This is precisely the expression for the anomalous velocity operator stemming from the leading order relativistic corrections in vacuum. It was conjectured in Appendix A of Ref.~\cite{adams1959energy}. Several later works~\cite{Chang_2008, vignale2010ten, PhysRevB.81.125332, RevModPhys.82.1959,Xiao_2018} have also mentioned this term.

\section{Relativistic corrections to the St\v reda's formula}

The relativistic (r) Kubo-Bastin formula can be derived in the Dirac theory by using standard time-dependent perturbation formalism. No matrix projections are needed for this. Hence no ambiguities appear. We can also assume that the unitary transformation $U$ has already been applied to the derived formula so that it is expressed as
\begin{multline}
    \label{r_Bastin}
    \sigma^{\text{r}}_{\alpha\beta}=\frac{e^2\hbar}{2\pi\Omega}\tr\!\int \!d\varepsilon\,f_\varepsilon\left[G_{U,-}^{\text{(D)}}\!-G_{U,+}^{\text{(D)}}\right] v_{U,\alpha}^{\text{(D)}} \left[G_{U,+}^{\text{(D)}}\right]' \!v_{U,\beta}^{\text{(D)}}\\+\text{c.c.}
\end{multline}
Eq.~\eqref{r_Bastin} has precisely the same form as Eq.~(\ref{nr_Bastin}) does, the only difference is the notations. 
The velocity operators here do not contain the vector potential corresponding to the electric field because the latter is already included in Eq.~(\ref{r_Bastin}) through the diamagnetic term.

Integration in Eq.~(\ref{r_Bastin}) is performed over the entire real line. We can split this single integration domain into two disjoint domains: one that corresponds to positrons and another one that corresponds to electrons. The two integrals that we obtain from Eq.~(\ref{r_Bastin}) by performing such a procedure determine Kubo-Bastin formulas for positrons and electrons, respectively. If the electron and positron branches are well separated, each of these two Kubo-Bastin formulas can be formulated exclusively in terms of projections of the involved operators onto the corresponding branch. Below, we derive such a ``projected'' Kubo-Bastin formula for electrons. For this we also assume that the Fermi level lies in the spectrum of $H^{\text{(el)}}$~\footnote{In fact, we assume that the Fermi level lies above the lower bound of the spectrum of $H^{\text{(el)}}$.} and that the temperature of the system is sufficiently small as compared to $m c^2$. Under these assumptions, the Kubo-Bastin formula for positrons simply provides a constant contribution to the total conductivity tensor. For brevity, we do not write it explicitly anywhere below.

Let us analyze the Green's functions that enter Eq.~(\ref{r_Bastin}), $G_{U,\pm}^{\text{(D)}}=(\varepsilon\pm i0-H_U^{\text{(D)}})^{-1}$. We have already shown that $H_U^{\text{(D)}}$ can be assumed diagonal with any desired accuracy. Therefore its inverse is diagonal too. Moreover, the positronic component of the latter is the same for the retarded and advanced Green's functions (we consider only the Kubo-Bastin formula for electrons). Hence, from Eq.~(\ref{HeHp}) one immediately derives
\begin{subequations}
\label{GeGp}
\begin{gather}
G_{U,\pm}^{\text{(D)}}=
\begin{pmatrix}
    G_{\pm}^{\text{(el)}} & 0\\
    0& G^{\text{(p)}}
\end{pmatrix},\\
G_{\pm}^{\text{(el)}}=(\varepsilon\pm i0-H^{\text{(el)}})^{-1},\quad
G^{\text{(p)}}=(\varepsilon-H^{\text{(p)}})^{-1}.
\end{gather}
\end{subequations}

Next we substitute Eqs.~(\ref{rv}) and (\ref{GeGp}) in Eq.~(\ref{r_Bastin}) and apply the matrix trace. This leads us to
\begin{multline}
\label{el_Bastin}
\sigma^{\text{r}}_{\alpha\beta}=\frac{e^2\hbar}{2\pi\Omega}\mtr\int d\varepsilon\,f_\varepsilon\Bigl(\bigl[G_{-}^{\text{(el)}}-G_{+}^{\text{(el)}}\bigr] v_{\alpha}^{\text{(el)}} \bigl[G_{+}^{\text{(el)}}\bigr]' v_{\beta}^{\text{(el)}}\\
+\bigl[G_{-}^{\text{(el)}}-G_{+}^{\text{(el)}}\bigr] \delta v_{\alpha} \bigl[G^{\text{(p)}}\bigr]' \delta v_{\beta}^\dagger\Bigr)+\text{c.c.},
\end{multline}
where $\mtr$ represents the operator trace over $2$-component spinors. The first term of the integrand above generalizes the nonrelativistic Kubo-Bastin formula of Eq.~(\ref{nr_Bastin}), the second term requires further treatment.

In Eq.~(\ref{vDU}) we have established a relation between $\delta \bb r$ and $\delta \bb v$. In terms of the Green's functions, it is reformulated as
\be
G_{\pm}^{\text{(el)}} \delta r_\alpha=\delta r_\alpha G^{\text{(p)}}-i\hbar\, G_{\pm}^{\text{(el)}} \delta v_\alpha G^{\text{(p)}}.
\e
Employing this result, its conjugate, and $\bigl[G^{\text{(p)}}\bigr]'=-\bigl[G^{\text{(p)}}\bigr]^2$, we obtain for the second term of the integrand in Eq.~(\ref{el_Bastin}):
\begin{multline}
\label{dvGdvG}
\mtr \Bigl(\bigl[G_{-}^{\text{(el)}}-G_{+}^{\text{(el)}}\bigr] \delta v_{\alpha} \bigl[G^{\text{(p)}}\bigr]' \delta v_{\beta}^\dagger\Bigr)+\text{c.c.}=\\-\frac{1}{\hbar^2}\mtr\Bigl(\bigl[\delta r_{\alpha}\delta r_{\beta}^\dagger-\delta r_{\beta}\delta r_{\alpha}^\dagger\bigr]\bigl[G_{-}^{\text{(el)}}-G_{+}^{\text{(el)}}\bigr]\Bigr).
\end{multline}
From the commutation relation $[r_{\alpha}^{\text{(D)}},r_{\beta}^{\text{(D)}}]=[r_{U,\alpha}^{\text{(D)}},r_{U,\beta}^{\text{(D)}}]=0$, one can also deduce
\be
\label{drdr}
\delta r_{\alpha}\delta r_{\beta}^\dagger-\delta r_{\beta}\delta r_{\alpha}^\dagger=-[r_{\alpha}^{\text{(el)}},r_{\beta}^{\text{(el)}}].
\e
We substitute Eqs.~(\ref{dvGdvG}) and (\ref{drdr}) in Eq.~(\ref{el_Bastin}) and apply the same procedure that was used to derive Eqs.~(\ref{nr_Streda}) from Eq.~(\ref{nr_Bastin}). This allows us to finally express the total dc conductivity tensor as $\sigma_{\alpha\beta}=\sigma^{\text{I}}_{\alpha\beta}+\sigma^{\text{II}}_{\alpha\beta}+\sigma^{\text{III}}_{\alpha\beta}$, where
\begin{subequations}
\label{el_Streda_final}
\begin{gather}
\label{el_Streda_1_final}
\sigma^{\text{I}}_{\alpha\beta}=\frac{e^2\hbar}{4\pi\Omega}\mtr\int d\varepsilon\,f'_\varepsilon\bigl[G_{+}^{\text{(el)}}-G_{-}^{\text{(el)}}\bigr] v_{\alpha}^{\text{(el)}} G_{+}^{\text{(el)}} v_{\beta}^{\text{(el)}}+\text{c.c.},
\\
\begin{multlined}
\label{el_Streda_2_final}
\sigma^{\text{II}}_{\alpha\beta}=\frac{e^2\hbar}{4\pi\Omega}\mtr\int d\varepsilon\,f_\varepsilon
\bigl[G_-^{\text{(el)}} v_\alpha^{\text{(el)}} G_-^{\text{(el)}} v_\beta^{\text{(el)}} G_-^{\text{(el)}}\phantom{df\,\,\,}\\ - G_-^{\text{(el)}} v_\beta^{\text{(el)}} G_-^{\text{(el)}} v_\alpha^{\text{(el)}} G_-^{\text{(el)}}\bigr]+\text{c.c.},
\end{multlined}
\\
\label{el_Streda_3_final}
\sigma^{\text{III}}_{\alpha\beta}=\frac{e^2}{2\pi\hbar\Omega}\mtr\int d\varepsilon\,f_\varepsilon\bigl[r_{\alpha}^{\text{(el)}},r_{\beta}^{\text{(el)}}\bigr]\bigl[G_{-}^{\text{(el)}}-G_{+}^{\text{(el)}}\bigr],\phantom{d}
\end{gather}
\end{subequations}
and we refrained from using additional letter superscripts to denote the tensors' elements.

Eqs.~(\ref{el_Streda_final}) provide the St\v reda's formula for electrons that consistently takes into account all effects of SOC. The first two terms, $\sigma^{\text{I}}_{\alpha\beta}$, $\sigma^{\text{II}}_{\alpha\beta}$, correspond to the conventional nonrelativistic Eqs.~(\ref{nr_Streda}), albeit with modified velocities. The third term, $\sigma^{\text{III}}_{\alpha\beta}$, to the best of our knowledge has been missing from the St\v reda's formula so far. To be more precise, it was never derived as a contribution that should be added \textit{to} Eqs.~(\ref{nr_Streda}). Ref.~\cite{PhysRevB.102.085113} derived it \textit{from} Eqs.~(\ref{nr_Streda}) and argued that it was the only intrinsic antisymmetric contribution to the dc conductivity tensor. This is however not the case because another important term was lost during that derivation. Moreover, as we have explained, Eqs.~(\ref{nr_Streda}) should not be used as a starting point for analysis of systems with SOC.

Contributions to transport effects from noncommuting coordinates were also derived in Ref.~\cite{Xiao_2018}. The same work presented a version of the Kubo formula for SHE that takes into account the anomalous velocity. Both results were obtained by adding relativistic corrections to the velocity and position operators in the effective (``projected'') model for electrons. However, the electron-positron matrix elements of the same operators were not taken into account in that work. We note that the term $\sigma^{\text{III}}_{\alpha\beta}$ defined in Eq.~(\ref{el_Streda_3_final}) originates from precisely these matrix elements.

In the context of SHE, the anomalous velocity operator in effective models was also considered, using kinetic-type equations, in Refs.~\cite{PhysRevB.81.125332, PhysRevB.88.035316}.

We emphasize that each velocity operator in Eqs.~(\ref{el_Streda_final}) contains the anomalous velocity and, in general, is not equal to $\partial H/\partial\bb p$. We also note that the position operator in Eq.~(\ref{el_Streda_3_final}) differs from $\bb r$, as it contains the Yafet term. Expansions of the velocity and position operators up to $1/c^2$ (for electrons) are provided by Eqs.~(\ref{rv_Pauli}). The relation between them that is valid in all orders reads
\be
\label{vrH}
\bb v^{\text{(el)}}=[\bb r^{\text{(el)}},H^{\text{(el)}}]/i\hbar.
\e
Eqs.~(\ref{el_Streda_final}) is the first central result of this paper~\footnote{When, in a crystal, $\bb r$, $\bb v$, $H$ are projected to a finite set of bands, it can happen that, for projected operators, $\bb v\neq[\bb r, H]/(i\hbar)$. In this case, Streda's formula that we derived is still valid, but one should choose such an operator for $\bb r^{\text{(D)}}$ that forms a canonical pair with momentum~\cite{ado2023position}.}.

\section{Anomalous Hall effect in the Rashba model}
We have developed a scheme suitable for derivation of the St\v reda's formula for Hamiltonians with a block structure where one of the blocks needs to be removed. Our scheme can be applied to crystals, particularly to narrow gap semiconductors, where spin-orbit effects are often much stronger than in vacuum~\cite{PhysRevLett.95.166605, sinova2008theory, vignale2010ten}. A widely used description for such systems is provided by the 8-band Kane model~\cite{KANE1957249}. It describes, in particular, electrons in the conduction band of semiconductors with zinc blende symmetry (e.g., GaAs, InSb). In such systems, the effective SOC originates from the interband matrix elements between conduction and valence bands~\cite{nozieres1973simple, sinova2008theory}. The latter are analogous to the matrix elements denoted by $h$ in Eq.~(\ref{h}). One can eliminate them and then derive the St\v reda's formula, Eqs.~(\ref{el_Streda_final}), just like we did for the Dirac system.

For a sample confined along one spatial direction by an asymmetric potential, the conduction electron Hamiltonian in the Kane model reduces to the famous two-dimensional Rashba (R) Hamiltonian~\cite{YuABychkov_1984} (if the bulk inversion asymmetry is weak)~\cite{RevModPhys.79.1217, kronmuller2007handbook, sinova2008theory}. In the presence of the internal magnetization in the $z$-direction, the latter reads
\be
\label{RASHBA}
H^{\text{(R)}}=\frac{\bb \pi^2}{2m^*}+\alpha_{\text{R}}[\bb \pi\times\bb\sigma]_z + M\sigma_z,
\e
where $m^*$ represents the effective mass, $M$ describes magnetization (or the exchange field), and $\alpha_{\text{R}}=\lambda\langle \nabla_z V_{\text{conf}}\rangle$ is proportional to the averaged $z$-derivative of the confining potential (we assume that the confinement is along $z$). The proportionality coefficient $\lambda$ is the analog of the relativistic prefactor $\hbar/4m^2c^2$ in front of the last term in Eq.~(\ref{He}). The analog of the $\bb\pi^4$ term is disregarded in $H^{\text{(R)}}$, while the confining potential itself and its second derivative correspond to a constant energy shift. Here we ignore ``relativistic'' corrections associated with the term $M\sigma_z$. However, they can also be important.

The model of Eq.~(\ref{RASHBA}) is particularly famous for predicting massive cancellations of SOC-related effects~\cite{PhysRevB.70.041303, PhysRevB.95.094401, PhysRevLett.97.046604}. The absence of dc AHE in the metallic regime of this model (in the noncrossing approximation, for Gaussian white-noise disorder) has been a subject of a very extensive debate~\cite{PhysRevLett.97.046604, PhysRevLett.99.066604, Kato_2007, PhysRevB.76.235312, PhysRevLett.100.236602, PhysRevB.79.195129, RevModPhys.82.1539, PhysRevLett.117.046601, CULCER2022}. We argue that the cancellation of the dc AHE in the Rashba model does not occur if correctly defined velocity (and position) operators are used. Instead of the operator $\bb v=\bb\pi/m^*-\alpha_{\text{R}}[\bb e_z\times\bb\sigma]$, which was used before, one should be using the operator analogous~\footnote{We disregard nonparabolic corrections to the velocity operator. They correspond to nonparabolic corrections to the Hamiltonian that we have already neglected.} to that of Eq.~(\ref{ve})~\footnote{In fact, this is also a simplification. The correct expression for the velocity operator in the Kane model (projected to the conduction band) includes an additional prefactor $\gamma\neq 0$ in front of the last term on the right-hand side of Eq.~(\ref{vR}). As a result, the expression for $\sigma^{\text{I}}_{xy}$ in Eq.~(\ref{sigmaR12a}) is multiplied by $\gamma^2$. The prefactor $\gamma$ is a function of the parameters of the Kane model. More detailed analysis of this model is presented in Ref.~\cite{ado2023position}}:
\be
\label{vR}
\bb v^{\text{(R)}}=\frac{\bb\pi}{m^*}-2\alpha_{\text{R}}[\bb e_z\times\bb\sigma]=\frac{\partial H^{\text{(R)}}}{\partial\bb\pi}-\alpha_{\text{R}}[\bb e_z\times\bb\sigma],
\e
where the last term is the anomalous velocity. 

After setting $\bb A=0$ in Eq.~(\ref{vR}) we can go along the lines of, e.g., Refs.~\cite{PhysRevB.76.235312, PhysRevB.101.085405}, and from Eqs.~(\ref{el_Streda_1_final}),~(\ref{el_Streda_2_final}) find for the Rashba model:
\be
\label{sigmaR12a}
\sigma^{\text{I}}_{xy}=
-\frac{e^2}{2\pi\hbar}\frac{\delta_{\text{SO}}\left(1+2\mu_{M}\delta_{\text{SO}}\right)}{\left(1+\mu_M\delta_{\text{SO}}\right)^2},
\qquad \sigma^{\text{II}}_{xy}=0,
\e
where $\delta_{\text{SO}}=\alpha_{\text{R}}^2m^*/M$, $\mu_{M}=\mu/M$, and $\mu$ is the chemical potential~\footnote{In Eq.~(\ref{sigmaR12a}), $\sigma^{\text{I}}_{xy}$ is given by a contribution quadratic with respect to $\alpha_{\text{R}}$. In principle, there is no guarantee that it is the only contribution to $\sigma^{\text{I}}_{xy}$ in this order.}. This result assumes the following conditions: the noncrossing approximation, Gaussian white-noise disorder, zero temperature, and metallic regime (both bands partly filled). Detailed explanations of these conditions' meanings can be found in Refs.~\cite{RevModPhys.82.1539,PhysRevB.76.235312, PhysRevB.101.085405}. We note that, in the derivation of Eq.~(\ref{sigmaR12a}), we disregarded relativistic corrections from the disorder potential. We also do not compute constant contributions to conductivity from the fully occupied lower bands of the Kane model.

In addition to the already finite $\sigma^{\text{I}}_{xy}$ in the Rashba model, we find a completely new contribution to the anomalous Hall dc conductivity from Eq.~(\ref{el_Streda_3_final}). Indeed, we have to assume that the coordinates are defined as
\be
\bb r^{\text{(R)}}=\bb r+\lambda[\bb\pi\times\bb\sigma]
\e
and thus do not commute: $[x^{\text{(R)}},y^{\text{(R)}}]=-2 i\lambda \hbar \sigma_z$, where we have disregarded magnetic fields. Hence, $\sigma^{\text{III}}_{xy}$ is proportional to the equilibrium spin polarization in the $z$-direction. This quantity was computed in a different context in Appendix~E of Ref.~\cite{PhysRevB.101.085405}. Using that result, we find from Eq.~(\ref{el_Streda_3_final}), in the metallic regime,
\be
\label{s2bgen}
\sigma^{\text{III}}_{xy}=-\frac{e^2}{\pi\hbar}\frac{2\lambda M m^*}{\hbar}.
\e

In the Kane model, $\lambda/\hbar=\delta g/4m\epsilon$, where $\epsilon$ is of the order of the band gap, $\delta g$ is the effective \mbox{$g$-factor} minus the bare \mbox{$g$-factor}, and $m$ is the free electron mass~\cite{nozieres1973simple, ado2023position}. By substituting this relation into Eq.~(\ref{s2bgen}), we obtain an expression for $\sigma^{\text{III}}_{xy}$ with a more transparent  meaning:
\begin{equation}
\label{s2bDR}
\sigma^{\text{III}}_{xy}=-\frac{e^2}{2\pi\hbar}\frac{M}{\epsilon}\frac{m^*\delta g}{m}.
\end{equation}
Up to a numerical prefactor, this intrinsic contribution to the anomalous Hall dc conductivity equals the ratio between the magnetization-induced spin splitting and the characteristic band gap energy. This contribution survives even without the Rashba coupling as long as $\lambda$ is finite. Indeed, in the latter case, relativistic corrections to the Hamiltonian $H^{\text{(R)}}$ and to the velocity operator vanish. However the spin-dependent part of the position operator still provides finite $\sigma^{\text{III}}_{xy}$ proportional to the relativistic parameter~$\lambda$.

Eqs.~(\ref{sigmaR12a}) and (\ref{s2bDR}) constitute our second central result.

\section{Discussion}
The St\v reda's formula that we derived, Eqs.~(\ref{el_Streda_final}), has two differences in comparison with its nonrelativistic (or fully relativistic~\footnote{In both the nonrelativistic and the fully relativistic theories, the system can be described by a Hamiltonian $H$ that acts in a certain Hilbert space. The velocity operator equals $\partial H/\partial\bb p$ in each of these two cases and acts in the same Hilbert space. Using this as a basis, one can employ the standard time-dependent perturbation formalism to derive the St\v reda's formula of Eqs.~(\ref{nr_Streda}) which, therefore, would be valid in both theories. However, when projections to subbands are involved, this scheme cannot be applied without a proper change, as we have shown in this work.}) version given by Eqs.~(\ref{nr_Streda}). The first difference is the deviation of the velocity operators from $\partial H/\partial\bb p$. It roots in the fact that the unitary transformation used to project the total Hamiltonian to a band (or a set of bands) depends on the canonical momentum~$\bb p$. Therefore, in the projected theory, the meaning of the operator $\partial/\partial \bb p$ changes. The second difference is the novel term $\sigma^{\text{III}}_{\alpha\beta}$ determined by the ``electron-positron'' matrix elements of the velocity operator.

Depending on the system, it may happen that only one of the two differences is present. For example, in the absence of the asymmetric potential in the Kane model, the corresponding anomalous velocity vanishes. The same happens in most of the models describing topological insulators.

If one does not perform any band projections, the differences do not occur at all. Thus, for example, the St\v reda's formula of Eqs.~(\ref{nr_Streda}) remains a valid tool for a fully relativistic computation of AHE~\cite{PhysRevLett.105.266604}. The conventional Berry-phase theory of AHE~\cite{RevModPhys.82.1959} is also justified, if the full matrix structure of the Hamiltonian is considered. For effective models, however, this theory has to be modified.

Finally, it should be noted that the importance of the interband components of the velocity operator was understood in several works on AHE, for two-level systems~\cite{PhysRevB.75.045315}. Computation of all Feynman diagrams involving such components represents a particular case of the analysis of $\sigma^{\text{III}}$.

\section{Conclusions}

Starting from the Dirac theory, we have derived the Kubo formula for the dc conductivity tensor (or the St\v reda's formula) of electrons in the presence of SOC. Each velocity operator in this formula contains the anomalous part. We also found an additional contribution to the Hall conductivity from noncommuting coordinates. Our scheme can be used to derive the St\v reda's formula for other Hamiltonians with a block structure. We have shown that AHE does not vanish in the Rashba model. We argue that Kubo formulas for other SOC-related effects should also be revised.

\begin{acknowledgments}
We are grateful to Misha Katsnelson, Koen Reijnders, and Achille Mauri for informative discussions. We thank Victoria Meshchaninova for designing the illustration. We thank Cong Xiao and Peter Oppeneer for valuable comments on the first version of the manuscript. I.\ A.\ A. and R.\ A.\ D. have received funding from the European Research Council (ERC) under the European Union’s Horizon 2020 research and innovation programme (Grant No.~725509). The Research Council of Norway (RCN) supported A.\ B. through its Centres of Excellence funding scheme, project number 262633, ``QuSpin'', and RCN project number 323766. M.\ T. has received funding from the European Union’s Horizon 2020 research and innovation program under the Marie Sk\l{}odowska-Curie grant agreement No 873028.
\end{acknowledgments}


\bibliography{Bib}

\end{document}